\documentclass[11pt,preprint]{aastex}

\received{}
\revised{}
\accepted{}
\ccc{}
\cpright{}{}

\slugcomment{}
\shorttitle{C and N on the 47 Tuc Main Sequence}
\shortauthors{Briley et al.}

\newcommand{\kms}{~km~s$^{-1}$}

\newcommand{\teff}{$T_{eff}$}
\newcommand{\grav}{log($g$)}

\begin{document}

\title{On the C and N Abundances of 47 Tucanae Main Sequence Stars\altaffilmark{1}}

\author{Michael M. Briley\altaffilmark{2}, Daniel Harbeck\altaffilmark{3},
Graeme H. Smith\altaffilmark{4}, and Eva K. Grebel\altaffilmark{5}}

\altaffiltext{1}{Based on observations made with ESO telescopes at the Paranal Observatory 
under program ID 67.D-0153 and the MPG/ESO 2.2 m telescope at La Silla 
Observatory}

\altaffiltext{2}{University of Wisconsin Oshkosh, 800 Algoma Blvd., Oshkosh,
Wisconsin 54901 (mike@maxwell.phys.uwosh.edu)}

\altaffiltext{3}{University of Wisconsin-Madison, 475 North Charter Street, Madison,
WI 53706 (harbeck@astro.wisc.edu)}

\altaffiltext{4}{UCO/Lick Observatory, University of California, Santa Cruz, 1156 High
Street, Santa Cruz, CA 95064 (graeme@ucolick.org)}

\altaffiltext{5}{Astronomical Institute of the University of Basel, Venusstrasse 7,
CH-4102 Binningen, Switzerland (grebel@astro.unibas.ch)}

\begin{abstract}
We report the results of an analysis of CN and CH band strengths among a large
sample of 47 Tucanae main-sequence and turn-off stars presented earlier by Harbeck
et al.
The resulting C and N abundances derived from synthetic spectra demonstrate:
1) A strongly anti-correlated relationship between [C/Fe] and [N/Fe] with the CN-strong
stars exhibiting depleted carbon and enhanced nitrogen.
2) The abundances of both elements 
agree remarkably well with those found among the evolved red giants of the cluster
implying little change in surface abundances from at least $M_{V} \approx +6.5$
to the tip of the red giant branch.
3) The pattern of C-depletions and N-enhancements are quite similar to those
seen among the turn-off stars of
M71, a cluster of almost identical metallicity but lower central concentration
and escape velocity.
At the same time, similar if not smaller N-enhancements and larger
C-depletions are evident among
like stars in the more metal-poor cluster M5.

We interpret these results, as did Harbeck et al., as evidence of the operation
of some pollution/accretion event early in the cluster history --- the most
likely source being AGB ejecta.
However, the present results rule out simple surface pollution and
suggest that a substantial fraction of the present stars' masses must 
be involved.

\end{abstract}


\section{Introduction}

It has been more than three decades since the first reports of star-to-star
differences in light element abundances among members 
of Galactic globular clusters
\citep{1971Obs....91..223O}.
That these intra-cluster inhomogeneities are confined to C, N, O, and often
Na, Al, and Mg, suggests an origin 
involving element production via
the CNO-cycle and associated proton-addition reactions.
Two likely sites for this nucleosynthesis exist:
One, within the presently evolving cluster stars
during red giant branch (RGB) ascent, where conditions above the H-burning
shell may allow material exposed to partial CN(O)-processing to be mixed to the
stellar surface (i.e., a deep dredge-up mechanism).
A second possible site lies in the asymptotic giant branch (AGB) phase of more
massive cluster members (now white dwarfs), whose ejecta were incorporated
into the present stars.

Unfortunately, the signatures of both sources appear remarkably similar and
observing the presence of CN(O)-cycle material in the photosphere of
an evolved cluster RGB star is not a sufficient discriminant as to its origin.
This has been made abundantly clear by the detection of O-Na and Mg-Al
anti-correlations among NGC 6752 main-sequence turn-off (MSTO)
and subgiant branch (SGB) stars by \citet{2001A&A...369...87G}.
The two possibilities may however be disentangled by exploring a span in
luminosity (evolutionary state) from the RGB tip to the main-sequence (MS).
Abundance changes with position on the RGB signal the operation of internal
processes, while composition patterns
present among MS stars are in place before RGB
ascent and the onset of deep mixing.

As telescope apertures and instrumentation have improved, such tests have
become increasingly feasible and data now exist on abundance variations
among low luminosity stars of a number of clusters
\citep*[47 Tuc, M71, M5, NGC 6752, and M13, see][for recent studies]
{1998MNRAS.298..601C, 1999AJ....117.2434C, 2001A&A...369...87G,
2002ApJ...579L..17B, 2002AJ....123.2525C, 2002ocuw.conf..119C, 2003AJ....125..197H}.
Perhaps not so surprisingly, we find ourselves now facing a wealth of evidence
that suggests not one origin or the other but rather both.
The progressive depletions in [C/Fe] with decreasing M$_V$ seen among the
stars of the more metal-poor clusters \citep[e.g. M92, M15, NGC 6397, see]
[and references therein]{2001PASP..113..326B}, the
drop in Li abundances at the luminosity function ``bump'' observed by
\citet{2002A&A...385L..14G} in NGC 6752, and the change in isotopic C
abundances with luminosity in NGC 6528 and M4 
found by \citet{2003ApJ...585L..45S}
provide dramatic evidence of a deep dredge-up mechanism.
Yet also, significant star-to-star differences in C and N have been found among
the faintest stars of every cluster studied to the level of the SGB or below.
Most recently, \citet{2002ApJ...579L..17B} documented substantial [C/Fe]
differences in a sample of M13 SGB stars which appeared to be subsequently
modified during RGB ascent.
Thus it would seem that in order to fully quantify the effects of deep mixing
one must know the compositions of the stars before they reach the RGB.

The process(es) which set the compositions of the cluster pre-RGB stars presumably
operated early in the cluster histories.
That large fractions of the members of every cluster studied to date show
significant differences of at least C and N imply this was not the result
of some unique condition among one or two clusters and it involved a non-trivial
portion of each cluster's mass.
As has been discussed by
\citet*{2001AJ....122.2561B} -- the process appears not to be a simple
accretion event.
It is evident then that a complete understanding of the origin of
abundance inhomogeneities in globular clusters must take account of
the pattern of light element abundances among the main sequence stars in
these objects.
With this in mind, we present a C and N abundance analysis of MS stars in
47 Tucanae, one of the most favorable globular clusters in the Milky Way
for the study of main sequence abundances.

We base our analysis on the data set obtained by \citet{2003AJ....125..197H},
who detected significant CN and CH variations as faint as 2.5~mag below the 
MSTO of 47 Tuc.
The occurrence of CN variations among these
low-mass stars favors accretion or primordial scenarios, since the capability
of such stars to maintain interior CNO-cycle burning is reduced by a
factor of at least ten compared to the main-sequence turn-off; thus such stars
are not expected to be able to modify their surface C and N abundances. 

\section{Observations, Indices, and Membership\label{section_spec}}

The sample selection, observations, and reduction of the 47 Tuc MS and
MSTO spectra used in this analysis are described in \citet{2003AJ....125..197H}.
To summarize, 115 program stars were selected from MPG/ESO 2.2 m images
of 47 Tuc to span a range in luminosity from the MSTO to $\approx 2.5$ mag
fainter.
Photometry of the stars was also extracted from these images.
The spectra were obtained at the ESO VLT using multi-slit masks with the Focal
Reducer and Low Dispersion Spectrograph 2 and cover from 3800
to 5500 \AA \ at R = 815.
The locations of the stars on the 47 Tuc color-magnitude diagram (CMD) are shown
in Figure \ref{fig_1}.

Membership was assessed in \citet{2003AJ....125..197H} based on radial velocity
\citep[unfortunately only $-18.7$\kms\ for 47 Tuc,][]{1996AJ....112.1487H}
and position in the CMD (within 0.025 mag of the cluster fiducial).
As a further test of membership, we have measured an MgH sensitive index
(MgH, with a feature bandpass from 5160 to 5195 \AA, and continuum bandpasses
from 5130 to 5160 and 5195 to 5225 \AA) which is plotted in Figure \ref{fig_2} (see also
Table \ref{tbl-1}).
Of the stars in the sample, an additional 11 have MgH band strengths inconsistent with
the other members of the cluster.
We will presume these stars to also be non-members of 47 Tuc and they will
not be included further.

Indices sensitive to absorption by CN at 3883 \AA \ and CH at 4300 \AA \ 
were measured from these spectra and are presented in \citet{2003AJ....125..197H}.
However, here we use the I(CH) index as defined in \citet{1999AJ....117.2428C} rather
than CH(4300) to aid in comparison with the low luminosity stars
of M71 \citep[see][]{1999AJ....117.2434C, 2001AJ....122..242B}.
The CN (S(3839)) and CH (I(CH)) indices of the stars in
the sample which are presumed members and further have measurable
CN, CH indices are given in Table \ref{tbl-1} and plotted in Figure \ref{fig_3},
where the CN-bimodality
and CN-CH anti-correlation noted by  \citet{2003AJ....125..197H} is quite evident.
The error bars shown are at the one-sigma level as determined from Poisson statistics
in the index bandpasses.

As noted in \citet{2002AJ....123.2525C} single-sided indices such as
S(3839) can suffer from zero-point offsets due to differences in slopes
caused by flat fielding, instrumental response, etc.
The observed spectra were therefore roughly flux calibrated before the
indices were measured by dividing by
a sensitivity function derived from a spectrum of LTT 377
taken during the same run.
Thus  the observed and model (see below) indices are assumed to share essentially
the same zero point  which should put the abundances on a scale
consistent with that used in previous analyses of other stars in this cluster.

\section{Model Atmospheres, Synthetic Spectra, and Resulting Abundances
\label{section_models}}

The approach we have taken is identical to that of \citet{2002AJ....123.2525C}:
establishing a \teff/\grav \ versus luminosity relation for a grid of model atmospheres
based on an isochrone fit, then adjusting the C and N abundances of corresponding synthetic
spectra to simultaneously fit the observed CN and CH band strengths.
The isochrone used was that of \citet{2001ApJ...556..322B} with an alpha-element
enhancement of 0.3 dex and an overall metallicity of $-0.83$.
An observed distance modulus of 13.33 and a reddening of 0.04 were
assumed following \citet*{2002A&A...395..481G}.
The fit of the isochrone to the region of the CMD of interest is shown in
Figure \ref{fig_1}.

For each of the program stars, a \teff \ and \grav \ was derived by transposing
its V magnitude onto the isochrone in Figure \ref{fig_1} and interpolating between grid points.
These values were then used as inputs to MARCS \citep{1975A&A....42..407G} for
the generation of model atmospheres.
The same overall metallicity ([Fe/H] = $-0.8$) as was used in previous studies
of more luminous 47 Tuc stars \citep[see][]{1997AJ....114.1051B} and of M71 MS/MSTO
stars \citep{2001AJ....122..242B} has been employed.
Synthetic spectra were computed from each model atmosphere for various C and N abundances
using the SSG program \citep*[][and references therein]
{1994MNRAS.268..771B} and the line-list of \citet{1995AJ....110.3035T}
\citep[see][for details]{2001AJ....122..242B}.
A microturbulent velocity of 2 \kms \ was assumed throughout the sample, and an
[O/Fe] of +0.45 was used \citep[see][for a discussion of the minor role O plays
in setting the CN/CH band strengths of 47 Tuc MS/MSTO stars]{1998MNRAS.298..601C}.
The synthetic spectra, calculated at steps of 0.05 \AA, were convolved with a
Gaussian of width 4.5 \AA \ to match the resolution of the observed spectra
and S(3839) and I(CH) indices measured from the result.
Sample synthetic indices appropriate to two sets of C and N abundances
over the luminosity range of the present sample are shown in Figure \ref{fig_3}.
Errors in the resulting abundances were calculated in a similar fashion using
indices adjusted by the $\pm1 \sigma$ uncertainties from Poisson statistics.

With two exceptions (stars 4554 and 6622 -- the brightest and faintest star in the
sample), all of the program star models yielded C and N abundances which 
simultaneously fit the observed indices to 0.005 or less (i.e., to within 0.01 dex
in [C/Fe] and [N/Fe], which is the limiting precision of the abundance inputs to the
synthesis program).
These values are shown in Table \ref{tbl-1}.
Under the present method of extracting abundances, it is the V photometry
of a star which is used to assign its \teff\ and \grav\ through fitting to the isochrone.
Thus, errors in the photometry as well as the assumed distance modulus will
lead to differences in the derived \teff, \grav, and resulting abundances.
Fortunately, with the exception of the MSTO region, the change in \teff, \grav\ 
with V is quite gradual.
As a test of this, we have recomputed the results for three pairs of CN-weak/strong
stars (one pair near the MSTO, one pair towards the luminous end of the MS sample,
and one pair near the faint end) assuming a distance modulus of 13.23 rather
than 13.33 (this is also equivalent to the change expected from a photometric
error of 0.1 in the brightness of an individual star).
These results are presented in Table \ref{tbl-2} where it can be seen that indeed, the
MS results are largely insensitive to errors in photometry/distance.

The resulting abundances (sans stars 4554 and 6622) are plotted in Figures
\ref{fig_3} and \ref{fig_4},
where a clear C versus N anti-correlation may be seen, especially in the case of
the CN-strong stars.
As is cautioned by \citet{2002AJ....123.2525C}, the method employed here can
naturally lead to such anti-correlations in the results (e.g., underestimating
[C/Fe] will naturally require larger [N/Fe] abundances to reproduce
a necessary CN band strength).
However, the indices of  Figure \ref{fig_3} lead us to conclude that the
anti-correlation of Figure \ref{fig_4} is likely quite real.

\section{Discussion
\label{section_results}}

We begin with a very brief summary of what is currently known about
abundance variations among low luminosity (MSTO/MS) stars in the
Galactic globular clusters, and in particular for 47 Tuc.
We consider here only such stars as they presumably reflect the
compositions of cluster members before the action of any deep mixing
mechanism:

1) Star-to-star abundance differences have been observed among the low
luminosity stars of 47 Tuc, NGC 6752, M5, M71, and M13 (see above
references).
In this sense, the intra-cluster light element abundance differences between low
luminosity stars appear ubiquitous among the Galactic globular clusters.
That similar patterns are not observed among metal-poor
field stars \citep*[e.g.,][]{2000A&A...356..238C} implies 
that some aspect of the cluster
environment plays a role.

2) The pattern of the MS/MSTO variations exhibits the same correlations
and anti-correlations expected from proton-capture nucleosynthesis, e.g.,
the C-N anti-correlation (see Figure \ref{fig_4}), the N-Na correlation of 47 Tuc \citep{1996Natur.383..604B},
and the O-Na, Mg-Al anti-correlations of NGC 6752 and M71 \citep{2001A&A...369...87G,2002AJ....123.3277R}.
It is widely accepted that the present-day cluster MS stars are incapable of
reaching the $\approx 10^8$ K temperatures required for these reactions,
which necessitates a source external to the present-day stars -- 
this consideration has lead to several so-called  ``primordial enrichment scenarios.''
The most often suggested site is intermediate-mass AGB stars undergoing hot
bottom burning and third dredge-up \citep[see][]{2001ApJ...550L..65V},
although difficulties such as the establishment of an O-Na anti-correlation
remain \citep[see for example][]{2003ApJ...590L..99D}.
Primordial scenarios include a number of possibilities as
discussed in \citet{1998MNRAS.298..601C}, among them: a pollution/accretion
process where already formed low-mass cluster stars acquire AGB ejecta
at a later date, or an extended period of star formation in which some stars
formed from gas already mixed with AGB ejecta.
Of course any viable scenario of this sort must not only explain the
star-to-star variations in light elements but also the remarkable consistency of
heavier (e.g. Fe-peak and $\alpha$-capture) elements within the clusters,
which can be a problem for some primordial theories (such as supernovae of
massive zero metallicity stars).

3) In the case of 47 Tuc and M71, little change in the range of C and N abundances
is observed from the tip of the RGB to the MSTO and below.
This is significant as the depth of the convective envelope varies significantly
over this interval (see Figure \ref{fig_1}) -- from a mass fraction of 0.068 at
0.65 $M_\odot$ (the lower
limit of the present sample) and 0.012 at 0.88 $M_\odot$ (our upper limit)
to a maximum fraction of 0.708 on the RGB at M$_V$ = +1.8 \citep{don03}.
In a pollution/accretion origin, any surface contamination would therefore
be diluted as the convective envelope deepens during RGB ascent, resulting
in increasing C and decreasing N abundances on RGB ascent.
That this is not observed then requires that a considerable
fraction ($\approx 70\%$ or more) of 
the mass of a CN-strong star include captured AGB ejecta.
As pointed out in \citet{2001AJ....122.2561B}, a reservoir of sufficient
C-poor/N-rich material may have been produced by more massive AGB
stars if the cluster initial
mass function is less steep than a Salpeter power law ($dN = m^{-\alpha} dm,$ 
i.e., $\alpha=1.5$ rather than 2.35), although without taking into account
several other factors such as efficiency with which the low-mass stars incorporate this
ejecta, $\alpha=1.5$ is at best an upper limit.

Note also that if a substantial fraction of cluster stars were to acquire
significant quantities of mass via the accretion of AGB ejecta,
one might expect a measurable change in the present day cluster
mass function.
Moreover, if the accretion process is indeed tied to the dense environment
of GCs as suggested by the lack of similar abundance variations among field main
sequence stars, a difference in the mass functions in low density environments
such as in the Local Group dwarf  spheroidal galaxies should be present. 
Yet the mass functions in the low-density environments of the dwarf
spheroidal galaxies Draco \citep{1998AJ....115..144G} and Ursa Minor
\citep{2002NewA....7..395W} appear largely indistinguishable from those
of at least the low-metallicity globular clusters,  in conflict with the expectations
from the AGB ejecta scenario.
However, if the accretion of CNO processed material happened during the epoch of
star formation, the mass function might remain unchanged.

The present results further highlight two issues in regard to such scenarios.
First,  the pattern of C and N abundances is the same for M71 and 47 Tuc -
see Figure \ref{fig_4} where the C and N abundances for a sample of
M71 MSTO/MS stars from \citet*{2001AJ....122..242B, briley03} are
plotted.
Yet the abundances found among similar stars in M5 by
\cite{2002AJ....123.2525C} show greater C depletions with similar if
not smaller N enhancements among the CN-strong stars (see also Figure \ref{fig_4}).
47 Tuc and M71 have nearly identical metallicities ([Fe/H] $\approx -0.8$),
and M5 is lower ([Fe/H] $\approx -1.2$).
The question is whether we are seeing differences in the initial C and N
abundances of the clusters, or a metallicity dependence in
the composition of the 
AGB-star ejecta presumed responsible for the
variations among the present-day MS stars.
Theory certainly predicts such a dependency
\citep[see for example][]{2001ApJ...550L..65V}, although one should
also therefore expect larger N over-abundances in M5 relative to
47 Tuc.
Clearly, without O, Na, Al, and/or Mg abundances among such stars,
and with only three clusters (two of the same metallicity) in the sample,
little more can be said.

Also note that despite the similarities in MSTO/MS compositions, the clusters
M71 and 47 Tuc have very different structural properties.
In the case of an origin for the CN-strong stars in material ejected from
AGB stars, one might expect a dependence on some aspect of
the cluster environment.
\citet{2002A&A...383..491T} discuss an accretion model in which
centrally concentrated clusters with high escape velocities can trap large
quantities of AGB-star ejecta, leading to the accretion of significant mass
by low mass members passing near the cluster center before the gas
is subsequently stripped during disk crossing \citep[as is apparently taking
place in NGC 6779,][]{2000MNRAS.316L...5H}.
In the case of 47 Tuc, an additional 0.8$M_\odot$ could be accumulated
by a 1$M_\odot$ star (0.4$M_\odot$ for M5) (see their Table 3).
Yet both the central concentration and escape velocity for M71 are
considerably lower \citep[$c$ = 1.15 versus 2.03, and $v_{esc, core}$ = 
16.7 versus 68.8 \kms,][respectively]{1996AJ....112.1487H, 
2002ApJ...568L..23G},
and this process should be far less efficient in M71 - to the point of being unable
to explain the $\approx 70\%$ or more
accreted material required for the \citet{2001AJ....122..242B} results.

\citet{2002PASP..114.1215S} recently revisited a correlation between the
fraction of CN-strong stars in a cluster and cluster  ellipticity  originally
noted by \citet{1987ApJ...313L..65N}.
Again, the  ellipticity  of M71 is 0.000 while that of 47 Tuc is 0.100 (see
Smith's Table 1).
However, also note the fraction of CN-strong stars is more than two times
greater in 47 Tuc compared to M71 ($r$ = 1.8 versus 0.7).
While this fraction is based on the distribution seen in larger surveys of RGB
stars, neither cluster exhibits evidence of significant composition changes
during RGB ascent (see above), thus this ratio also likely reflects the
fraction of contaminated stars on the MS.
Considering the similarities of such stars in 47 Tuc and M71,
it would then appear that we are seeing the action of the same process,
resulting in the same composition modifications -
just a smaller fraction of the cluster stars were involved in M71.
As pointed out by Smith, the CN-strong/ellipticity correlation
may be a selection effect -- the higher  ellipticity  clusters in the
sample tend also to have higher masses \citep[according to the compilation of][47
Tuc has both a considerably higher velocity dispersion and a higher 
inferred mass than M71]{1993sdgc.proc..357P}.
Whether the difference between M71 and 47 Tuc  is the result of a smaller
central reservoir of AGB ejecta or less mass involved in the tail end of star
formation (and how the necessary gas could survive being blown out by
supernovae), or some aspect of the cluster dynamics, remains to be seen.

\acknowledgments

We wish to thank Roger Bell for the use of the SSG program,
Don VandenBerg for providing the convective envelope mass fractions,
and the anonymous referee for their helpful comments.
MMB acknowledges support from the National Science Foundation 
(under grant AST-0098489), the F. John Barlow professorship, and the UW
Oshkosh Faculty Development Program.
DH is supported in part through the McKinney postdoctoral fellowship at 
the University of Wisconsin-Madison. 
GHS gratefully acknowledges the support of NSF grant AST-0098453.

\clearpage

\begin{figure}
\epsscale{0.7}
\caption[Briley.fig1.eps]{
The CMD of 47 Tuc is plotted with the locations of the
program stars marked including the fiducial points of
\citet{1987PASP...99..739H} and the locations of the isochrone
models from \cite{2001ApJ...556..322B}.
HST photometry of stars near the core of 47 Tuc as provided by MMB
(as part of a forthcoming paper) is also plotted.
The numbers assigned to arrows indicate the initial mass of the model
in the isochrone and the mass fraction involved in the outer convective
envelope while at that point in its evolution, respectively \citep[from][]{don03}.
\label{fig_1}}
\end{figure}

\begin{figure}
\epsscale{0.7}
\caption[Briley.fig2.eps]{
The MgH indices as defined in the text are plotted versus
V. Stars excluded as likely non-members based on either
radial velocities, position on the CMD \citep{2003AJ....125..197H},
or MgH index are marked as
``x'' while CN-weak and CN-strong presumed members
are marked with open and closed circles (respectively).
\label{fig_2}}
\end{figure}

\begin{figure}
\epsscale{0.7}
\caption[Briley.fig3.eps]{
The CN and CH band indices (left upper and lower panels, respectively)
are plotted versus $V$ magnitude and $M_V$. The symbols are as
in Figure \ref{fig_2}. As pointed out by \citet{2003AJ....125..197H},
the distribution of CN band strengths appears bimodal, and a
general anti-correlation between CN and CH is apparent. Also
shown are two isoabundance lines computed from the isochrone
models for two different C and N abundances ([C/Fe]/[N/Fe]/[O/Fe]
are listed in parentheses). These are essentially the same set of
abundances that fit CN and CH bands among evolved 47 Tuc stars
\citep{1997AJ....114.1051B}, and the MSTO/RGB stars of M71
\citep{2001AJ....122..242B, 2001AJ....122.2561B} -- a cluster of
very similar metallicity.
The resulting N and C abundances are also plotted as a function
of luminosity (right upper and lower panels, respectively).
As expected, an overall anti-correlation between C and N exists.
However, no real change in the pattern or range of abundances
takes place from the MSTO to some 2.5 mags below.
\label{fig_3}}
\end{figure}

\begin{figure}
\epsscale{0.7}
\caption[Briley.fig4.eps]{
Resulting C and N abundances measured by matching CH and CN
band strengths are plotted against each other. (Left) Only the results
for the present sample of 47 Tuc MS stars are shown, where the
strong anti-correlation between C and N can be seen among the
CN-strong stars. In contrast, the CN-weak stars exhibit only a
modest range in [C/Fe].
(Right) The distribution of C and N among a
similar sample of M71 and M5 stars \citep[from][]{2002ApJ...579L..17B} is also plotted.
The general pattern (i.e., a C/N anti-correlation) is the same
between all three clusters, and M71 and 47 Tuc (two clusters of similar
metallicity) appear to be essentially indistinguishable.
In the case of the more metal-poor M5, the C depletions are more
extreme, yet the N enhancements are not.
\label{fig_4}}
\end{figure}

\clearpage

\renewcommand{\arraystretch}{.6}

\begin{deluxetable}{ccccccccccccc}
\tabletypesize{\scriptsize}
\tablecolumns{13}
\tablewidth{0pc}
\tablecaption{Observed Indices, Model Parameters, and Resulting Abundances
\label{tbl-1}}
\tablehead{
\colhead{ID}  & \colhead{V\tablenotemark{a}}  & \colhead{B$-$V\tablenotemark{a}}  &
\colhead{S(3839)} & \colhead{I(CH)} & \colhead{MgH\tablenotemark{b}} & \colhead{Member} & \colhead{T$_{eff}$} &
\colhead{log g} & \colhead{[C/Fe]} & \colhead{[N/Fe]} & \colhead{Model S(3839)} & \colhead{Model I(CH)}
}
\startdata
90 & 19.695 & 0.831 & -0.360 & 0.697 & 0.007 & Y & 5141 & 4.62 & 0.04 & 0.14 & -0.361 & 0.698 \\
94 & 18.675 & 0.601 & -0.062 & 0.657 & -0.086 & Y & 5709 & 4.50 & -0.31 & 1.33 & -0.060 & 0.657 \\
341 & 17.575 & 0.534 & -0.262 & 0.654 & -0.101 & Y & 5953 & 4.18 & -0.07 & 1.01 & -0.261 & 0.655 \\
596 & 19.675 & 0.782 & -0.319 & 0.694 & 0.041 & Y & 5153 & 4.62 & -0.01 & 0.30 & -0.320 & 0.694 \\
806 & 19.352 & 0.737 & -0.006 & 0.678 & -0.002 & Y & 5350 & 4.60 & -0.17 & 1.12 & -0.010 & 0.679 \\
977 & 18.825 & 0.650 & -0.107 & 0.678 & -0.048 & Y & 5634 & 4.52 & -0.12 & 0.98 & -0.109 & 0.676 \\
1152 & 19.335 & 0.765 & 0.444 & 0.622 & 0.077 & N & - & - & - & - & - & - \\
1173 & 17.790 & 0.554 & -0.392 & 0.661 & -0.108 & Y & 6024 & 4.29 & 0.07 & 0.55 & -0.391 & 0.661 \\
1304 & 17.665 & 0.564 & -0.222 & 0.644 & - & ? & 5975 & 4.22 & -0.21 & 1.27 & -0.222 & 0.645 \\
1329 & 18.401 & 0.627 & -0.355 & 0.671 & -0.070 & Y & 5824 & 4.44 & -0.06 & 0.54 & -0.355 & 0.671 \\
1336 & 17.921 & 0.584 & -0.161 & 0.638 & -0.082 & Y & 6008 & 4.33 & -0.31 & 1.53 & -0.161 & 0.638 \\
1351 & 20.375 & 0.931 & -0.393 & 0.666 & -0.096 & N & - & - & - & - & - & - \\
1462 & 18.265 & 0.551 & -0.150 & 0.635 & -0.090 & Y & 5871 & 4.41 & -0.44 & 1.48 & -0.149 & 0.640 \\
1498 & 17.642 & 0.521 & -0.250 & 0.620 & -0.155 & N & - & - & - & - & - & - \\
1587 & 18.095 & 0.600 & -0.387 & 0.669 & -0.082 & Y & 5936 & 4.37 & 0.06 & 0.45 & -0.387 & 0.670 \\
1675 & 18.555 & 0.601 & -0.549 & 0.679 & 0.145 & N & - & - & - & - & - & - \\
1833 & 19.835 & 0.798 & -0.364 & 0.696 & 0.052 & N & - & - & - & - & - & - \\
1992 & 18.815 & 0.633 & -0.258 & 0.689 & -0.049 & Y & 5639 & 4.52 & 0.02 & 0.54 & -0.260 & 0.689 \\
2060 & 18.175 & 0.584 & -0.201 & 0.655 & -0.095 & Y & 5903 & 4.39 & -0.16 & 1.14 & -0.200 & 0.655 \\
2186 & 18.905 & 0.634 & -0.023 & 0.655 & -0.049 & Y & 5593 & 4.53 & -0.41 & 1.42 & -0.021 & 0.655 \\
2284 & 19.955 & 0.799 & 0.137 & 0.686 & 0.045 & Y & 4985 & 4.64 & 0.07 & 1.47 & 0.137 & 0.686 \\
2379 & 18.459 & 0.559 & -0.137 & 0.671 & -0.095 & Y & 5803 & 4.46 & -0.05 & 1.03 & -0.135 & 0.671 \\
2409 & 17.205 & 0.601 & -0.157 & 0.669 & -0.060 & N & - & - & - & - & - & - \\
2556 & 19.225 & 0.733 & -0.326 & 0.696 & 0.006 & Y & 5422 & 4.58 & -0.01 & 0.30 & -0.326 & 0.695 \\
2629 & 18.165 & 0.567 & -0.380 & 0.673 & -0.095 & Y & 5907 & 4.38 & 0.07 & 0.43 & -0.380 & 0.673 \\
2738 & 19.965 & 0.881 & 0.116 & 0.679 & 0.057 & Y & 4979 & 4.64 & -0.01 & 1.54 & 0.116 & 0.680 \\
2821 & 19.165 & 0.716 & 0.045 & 0.658 & -0.025 & Y & 5455 & 4.57 & -0.37 & 1.46 & 0.045 & 0.660 \\
2947 & 18.605 & 0.617 & -0.351 & 0.687 & -0.055 & Y & 5742 & 4.49 & 0.05 & 0.35 & -0.356 & 0.686 \\
3060 & 17.465 & 0.551 & -0.383 & 0.656 & -0.109 & Y & 6089 & 4.20 & 0.10 & 0.67 & -0.383 & 0.655 \\
3137 & 17.199 & 0.582 & -0.380 & 0.674 & -0.106 & Y & 5688 & 3.94 & -0.11 & 0.34 & -0.381 & 0.674 \\
3206 & 19.545 & 0.716 & -0.331 & 0.646 & -0.142 & N & - & - & - & - & - & - \\
3230 & 19.105 & 0.700 & -0.323 & 0.394 & 0.110 & N & - & - & - & - & - & - \\
3343 & 18.642 & 0.615 & -0.061 & 0.648 & -0.078 & Y & 5725 & 4.49 & -0.45 & 1.49 & -0.060 & 0.647 \\
3371 & 17.425 & 0.567 & -0.275 & 0.658 & -0.099 & Y & 6143 & 4.21 & 0.22 & 1.01 & -0.276 & 0.658 \\
3394 & 18.870 & 0.675 & -0.035 & 0.670 & -0.063 & Y & 5611 & 4.53 & -0.20 & 1.19 & -0.033 & 0.670 \\
3456 & 19.525 & 0.799 & -0.405 & 0.695 & 0.025 & Y & 5246 & 4.61 & -0.02 & 0.07 & -0.406 & 0.695 \\
3521 & 18.465 & 0.637 & -0.223 & 0.677 & -0.076 & Y & 5801 & 4.46 & 0.02 & 0.79 & -0.218 & 0.678 \\
3522 & 19.954 & 0.794 & 0.017 & 0.691 & 0.065 & Y & 4985 & 4.64 & 0.08 & 1.11 & 0.016 & 0.691 \\
3532 & 17.395 & 0.551 & -0.366 & 0.664 & -0.106 & Y & 6148 & 4.20 & 0.30 & 0.65 & -0.366 & 0.664 \\
3631 & 19.625 & 0.848 & -0.062 & 0.672 & 0.070 & Y & 5184 & 4.62 & -0.27 & 1.18 & -0.063 & 0.671 \\
3701 & 19.215 & 0.716 & 0.002 & 0.681 & -0.014 & Y & 5428 & 4.58 & -0.13 & 1.11 & 0.004 & 0.681 \\
3769 & 18.235 & 0.567 & -0.155 & 0.650 & -0.082 & Y & 5881 & 4.40 & -0.26 & 1.30 & -0.155 & 0.650 \\
3797 & 17.937 & 0.576 & -0.366 & 0.673 & -0.087 & Y & 6003 & 4.33 & 0.20 & 0.50 & -0.367 & 0.673 \\
3823 & 18.268 & 0.637 & -0.287 & 0.700 & -0.082 & Y & 5870 & 4.41 & 0.34 & 0.41 & -0.288 & 0.699 \\
3853 & 18.255 & 0.584 & -0.185 & 0.659 & -0.086 & Y & 5874 & 4.41 & -0.14 & 1.12 & -0.180 & 0.659 \\
3872 & 18.389 & 0.657 & -0.365 & 0.685 & -0.065 & Y & 5828 & 4.44 & 0.12 & 0.34 & -0.366 & 0.685 \\
3929 & 19.425 & 0.782 & -0.265 & 0.700 & 0.012 & Y & 5307 & 4.60 & 0.05 & 0.35 & -0.268 & 0.700 \\
3942 & 19.398 & 0.813 & -0.355 & 0.695 & 0.000 & Y & 5323 & 4.60 & -0.02 & 0.22 & -0.354 & 0.695 \\
4014 & 18.155 & 0.584 & -0.359 & 0.674 & -0.100 & Y & 5911 & 4.38 & 0.09 & 0.49 & -0.360 & 0.674 \\
4019 & 19.952 & 0.889 & -0.307 & 0.687 & 0.086 & N & - & - & - & - & - & - \\
4023 & 19.610 & 0.812 & 0.043 & 0.677 & - & ? & 5194 & 4.62 & -0.17 & 1.30 & 0.044 & 0.677 \\
4115 & 18.005 & 0.567 & -0.364 & 0.668 & -0.093 & Y & 5976 & 4.35 & 0.09 & 0.57 & -0.363 & 0.668 \\
4188 & 19.815 & 0.798 & -0.272 & 0.693 & 0.041 & Y & 5067 & 4.63 & -0.01 & 0.44 & -0.273 & 0.692 \\
4298 & 18.550 & 0.648 & -0.135 & 0.668 & -0.068 & Y & 5766 & 4.48 & -0.14 & 1.07 & -0.137 & 0.667 \\
4344 & 17.735 & 0.567 & -0.272 & 0.650 & -0.104 & Y & 6011 & 4.26 & -0.09 & 1.08 & -0.273 & 0.650 \\
4421 & 18.759 & 0.642 & -0.045 & 0.668 & -0.053 & Y & 5668 & 4.51 & -0.18 & 1.19 & -0.045 & 0.669 \\
4452 & 18.733 & 0.587 & -0.405 & 0.678 & - & ? & 5681 & 4.51 & -0.10 & 0.29 & -0.401 & 0.678 \\
4465 & 18.591 & 0.647 & 0.020 & 0.632 & - & ? & 5748 & 4.48 & -0.69 & 1.93 & 0.019 & 0.632 \\
4521 & 19.330 & 0.734 & 0.004 & 0.672 & 0.023 & Y & 5362 & 4.59 & -0.26 & 1.25 & 0.004 & 0.671 \\
4554 & 17.124 & 0.711 & -0.066 & 0.602 & - & ? & 4404 & 3.42 & -1.01 & 2.49 & -0.067 & 0.641 \\
4592 & 19.401 & 0.772 & -0.368 & 0.639 & -0.131 & N & - & - & - & - & - & - \\
4600 & 18.905 & 0.634 & -0.313 & 0.697 & -0.051 & Y & 5593 & 4.53 & 0.08 & 0.32 & -0.314 & 0.697 \\
4643 & 17.625 & 0.567 & -0.416 & 0.660 & -0.100 & Y & 5957 & 4.20 & -0.01 & 0.41 & -0.416 & 0.660 \\
4707 & 18.128 & 0.569 & -0.156 & 0.641 & -0.095 & Y & 5922 & 4.38 & -0.37 & 1.47 & -0.156 & 0.640 \\
4790 & 18.565 & 0.617 & -0.184 & 0.674 & -0.075 & Y & 5760 & 4.48 & -0.06 & 0.89 & -0.185 & 0.674 \\
4877 & 18.585 & 0.699 & -0.124 & 0.677 & -0.039 & Y & 5751 & 4.48 & -0.04 & 0.98 & -0.125 & 0.676 \\
4936 & 18.065 & 0.584 & -0.174 & 0.651 & -0.088 & Y & 5949 & 4.36 & -0.16 & 1.26 & -0.177 & 0.651 \\
5049 & 17.135 & 0.666 & -0.223 & 0.683 & -0.089 & Y & 4644 & 3.52 & -0.23 & 0.53 & -0.218 & 0.683 \\
5146 & 20.077 & 0.891 & -0.064 & 0.688 & - & ? & 4915 & 4.65 & 0.03 & 1.06 & -0.064 & 0.687 \\
5154 & 18.505 & 0.601 & -0.382 & 0.682 & -0.084 & Y & 5785 & 4.47 & 0.04 & 0.31 & -0.383 & 0.682 \\
5188 & 17.485 & 0.567 & -0.240 & 0.632 & -0.115 & Y & 6053 & 4.19 & -0.36 & 1.51 & -0.241 & 0.632 \\
5205 & 17.350 & 0.553 & -0.192 & 0.629 & -0.101 & Y & 6053 & 4.14 & -0.39 & 1.66 & -0.193 & 0.630 \\
5214 & 19.859 & 0.728 & -0.033 & 0.590 & -0.086 & N & - & - & - & - & - & - \\
5219 & 18.656 & 0.629 & -0.395 & 0.681 & -0.069 & Y & 5718 & 4.49 & -0.03 & 0.27 & -0.396 & 0.681 \\
5231 & 18.865 & 0.666 & 0.025 & 0.689 & 0.042 & N & - & - & - & - & - & - \\
5235 & 19.975 & 0.881 & -0.251 & 0.692 & 0.089 & N & - & - & - & - & - & - \\
5237 & 19.383 & 0.784 & -0.375 & 0.694 & 0.032 & Y & 5331 & 4.60 & -0.04 & 0.19 & -0.374 & 0.694 \\
5253 & 19.007 & 0.698 & 0.053 & 0.667 & -0.027 & Y & 5540 & 4.55 & -0.24 & 1.36 & 0.054 & 0.668 \\
5261 & 20.383 & 0.904 & -0.263 & 0.701 & -0.134 & N & - & - & - & - & - & - \\
5296 & 19.205 & 0.716 & -0.375 & 0.648 & -0.123 & N & - & - & - & - & - & - \\
5299 & 19.281 & 0.708 & -0.308 & 0.698 & 0.015 & Y & 5390 & 4.59 & 0.03 & 0.30 & -0.306 & 0.698 \\
5342 & 17.605 & 0.551 & -0.244 & 0.640 & -0.105 & Y & 5952 & 4.19 & -0.35 & 1.32 & -0.245 & 0.639 \\
5381 & 19.175 & 0.716 & -0.002 & 0.669 & -0.123 & N & - & - & - & - & - & - \\
5402 & 18.505 & 0.650 & -0.368 & 0.684 & - & ? & 5785 & 4.47 & 0.06 & 0.34 & -0.369 & 0.684 \\
5538 & 18.445 & 0.617 & -0.236 & 0.678 & -0.098 & Y & 5808 & 4.45 & 0.03 & 0.74 & -0.241 & 0.678 \\
5545 & 18.764 & 0.678 & -0.359 & 0.688 & -0.048 & Y & 5665 & 4.51 & 0.03 & 0.30 & -0.358 & 0.689 \\
5547 & 19.525 & 0.815 & 0.223 & 0.672 & 0.035 & Y & 5246 & 4.61 & -0.16 & 1.69 & 0.223 & 0.672 \\
5587 & 17.829 & 0.589 & -0.392 & 0.668 & -0.096 & Y & 6025 & 4.30 & 0.17 & 0.46 & -0.390 & 0.669 \\
5653 & 18.174 & 0.602 & -0.232 & 0.654 & -0.096 & Y & 5903 & 4.39 & -0.16 & 1.07 & -0.232 & 0.655 \\
5665 & 17.282 & 0.584 & -0.193 & 0.631 & -0.111 & Y & 5624 & 3.94 & -0.91 & 1.56 & -0.193 & 0.631 \\
5759 & 17.446 & 0.554 & -0.244 & 0.642 & - & ? & 6120 & 4.20 & -0.06 & 1.32 & -0.245 & 0.642 \\
5809 & 19.371 & 0.764 & -0.370 & 0.694 & 0.017 & Y & 5338 & 4.60 & -0.04 & 0.20 & -0.370 & 0.694 \\
5836 & 18.885 & 0.698 & -0.110 & 0.683 & -0.025 & Y & 5604 & 4.53 & -0.06 & 0.90 & -0.109 & 0.683 \\
5942 & 19.271 & 0.780 & 0.104 & 0.676 & 0.011 & Y & 5396 & 4.58 & -0.18 & 1.37 & 0.105 & 0.675 \\
6043 & 17.938 & 0.567 & -0.193 & 0.631 & -0.094 & Y & 6002 & 4.33 & -0.45 & 1.61 & -0.190 & 0.632 \\
6047 & 17.246 & 0.602 & -0.385 & 0.666 & -0.106 & Y & 5459 & 3.86 & -0.42 & 0.41 & -0.385 & 0.665 \\
6220 & 18.665 & 0.731 & -0.041 & 0.645 & -0.068 & Y & 5714 & 4.50 & -0.50 & 1.57 & -0.040 & 0.644 \\
6248 & 19.047 & 0.774 & -0.328 & 0.691 & -0.022 & Y & 5519 & 4.55 & -0.04 & 0.36 & -0.328 & 0.690 \\
6264 & 19.999 & 0.797 & -0.001 & 0.668 & 0.065 & Y & 4959 & 4.64 & -0.23 & 1.55 & 0.000 & 0.668 \\
6335 & 19.335 & 0.739 & -0.348 & 0.634 & -0.151 & N & - & - & - & - & - & - \\
6380 & 19.813 & 0.775 & 0.091 & 0.670 & 0.047 & Y & 5069 & 4.63 & -0.23 & 1.64 & 0.091 & 0.668 \\
6452 & 18.635 & 0.666 & -0.067 & 0.668 & -0.065 & Y & 5728 & 4.49 & -0.15 & 1.17 & -0.069 & 0.668 \\
6455 & 18.029 & 0.614 & -0.232 & 0.660 & -0.093 & Y & 5965 & 4.35 & -0.01 & 1.03 & -0.227 & 0.660 \\
6537 & 17.326 & 0.589 & -0.316 & 0.658 & - & ? & 5935 & 4.08 & -0.02 & 0.80 & -0.315 & 0.660 \\
6591 & 17.572 & 0.574 & -0.236 & 0.641 & -0.111 & Y & 5954 & 4.18 & -0.27 & 1.27 & -0.235 & 0.643 \\
6622 & 22.021 & 0.259 & -0.356 & 0.604 & -0.181 & Y & 3977 & 4.76 & 0.04 & 2.35 & -0.361 & 0.646 \\
6641 & 19.403 & 0.779 & 0.101 & 0.664 & 0.017 & Y & 5320 & 4.60 & -0.33 & 1.56 & 0.101 & 0.663 \\
6648 & 19.219 & 0.724 & 0.083 & 0.672 & -0.022 & Y & 5425 & 4.58 & -0.21 & 1.35 & 0.079 & 0.672 \\
6723 & 19.214 & 0.733 & 0.121 & 0.669 & -0.018 & Y & 5428 & 4.58 & -0.23 & 1.46 & 0.120 & 0.670
\enddata
\tablenotetext{a}{Note that the V, B$-$V values of Harbeck et al. (2003) include an extinction correction
of 0.128 and 0.04 (respectively). The values listed here are uncorrected.}
\tablenotetext{b}{Nine spectra did not extend redward enough for the MgH index to be measured. They are
assumed to be members based on the criteria of Harbeck (2003) and included in the subsequent analysis.
No modling was done for the presumed non-members.}
\end{deluxetable}

\clearpage

\begin{deluxetable}{rrrrrrrrrrrrrrrrrrrrrrr}
\rotate
\tabletypesize{\scriptsize}
\tablecolumns{22}
\tablewidth{0pt}
\tablecaption{Model Fits and Resulting Abundances, Including a Change in Distance Modulus of 0.10 mags.
\label{tbl-2}}
\tablehead{
\colhead{} & \colhead{} & \colhead{} & \colhead {} & \colhead {} &
\multicolumn{6}{c}{(M$-$m)$_V$ = 13.33} & \colhead {} & \multicolumn{6}{c}{(M$-$m)$_V$ = 13.23} &
\colhead{} & \multicolumn{4}{c}{Change}\\
\cline{6-11} \cline{13-18} \cline{20-23} \\
\colhead{Star} & \colhead{V} & \colhead{S(3839)} & \colhead{I(CH)} & \colhead {} & \colhead{\teff} &
\colhead{\grav} & \colhead{[C/Fe]} & \colhead{[N/Fe]} & \colhead{S(3839)} & \colhead{I(CH)} & \colhead {}  &
\colhead{\teff} & \colhead{\grav} & \colhead{[C/Fe]} & \colhead{[N/Fe]} & \colhead{S(3839)} & \colhead{I(CH)} &
\colhead {} & \colhead{\teff} & \colhead{\grav} & \colhead{[C/Fe]} & \colhead{[N/Fe]}
}
\startdata

5205  & 17.350 & -0.192 & 0.629 & & 6053 & 4.14 & -0.39 & 1.66 & -0.193 & 0.630 & & 6114 & 4.20 & -0.34 & 1.72 & -0.193 & 0.629 & &  61 &  0.06 &  0.05 &  0.06 \\
6537  & 17.326 & -0.316 & 0.658 & & 5935 & 4.08 & -0.02 & 0.80 & -0.315 & 0.660 & & 6143 & 4.21 &  0.21 & 0.91 & -0.314 & 0.657 & & 208 &  0.13 &  0.23 &  0.11 \\
 &  &  &  &  &  &  &  &  &  & &  &  &  &  & & &  &  &  &  &  &  \\
1336  & 17.921 & -0.161 & 0.638 & & 6008 & 4.33 & -0.31 & 1.53 & -0.161 & 0.638 & & 5969 & 4.35 & -0.35 & 1.51 & -0.163 & 0.638 & & -39 &  0.02 & -0.04 & -0.02 \\
3797  & 17.937 & -0.366 & 0.673 & & 6003 & 4.33 &  0.20 & 0.50 & -0.367 & 0.673 & & 5962 & 4.36 &  0.13 & 0.50 & -0.366 & 0.672 & & -41 &  0.03 & -0.07 &  0.00 \\
 &  &  &  &  &  &  &  &  & &  &  &  &  &  & & &  &  &  &  &  &  \\
6641  & 19.403 &  0.101 & 0.664 & & 5320 & 4.60 & -0.33 & 1.56 &  0.101 & 0.663 & & 5260 & 4.61 & -0.32 & 1.58 &  0.100 & 0.664 & & -60 &  0.01 &  0.01 &  0.02 \\
3929  & 19.425 & -0.265 & 0.700 & & 5307 & 4.60 &  0.05 & 0.35 & -0.268 & 0.700 & & 5246 & 4.61 &  0.05 & 0.35 & -0.266 & 0.700 & & -61 &  0.01 &  0.00 &  0.00
\enddata
\end{deluxetable}

\end{document}